\documentclass[aps,pra,twocolumn,showpacs,superscriptaddress,groupedaddress]{revtex4}
\usepackage{amsmath}
\usepackage{amsthm}
\usepackage{newlfont}
\usepackage{graphicx}
\usepackage{epstopdf}
\usepackage{appendix}
\usepackage{textcomp}
\usepackage{appendix}
\usepackage{multirow}
\usepackage{bbding}
\usepackage{pifont}
\usepackage{wasysym}
\usepackage{wrapfig}
\usepackage{placeins}

\begin{document}

\title{ Modulational instability in binary spin-orbit-coupled Bose-Einstein
condensates}
\author{Ishfaq Ahmad Bhat$^1$, T. Mithun$^1$, B. A. Malomed$^2$, and K.
Porsezian$^1$}
\address{$^1$Department of Physics, Pondicherry University, Puducherry
605014, India\\
$^2$Department of Physical Electronics, School of Electrical Engineering,
Faculty of Engineering, Tel Aviv University, Tel Aviv 69978, Israel}

\begin{abstract}
We study modulation instability (MI) of flat states in two-component
spin-orbit-coupled (SOC) Bose-Einstein condensates (BECs) in the framework
of coupled Gross-Pitaevskii equations for two components of the
pseudo-spinor wave function. The analysis is performed for equal densities
of the components. Effects of the interaction parameters, Rabi coupling and
SOC on the MI are investigated. In particular, the results demonstrates that
the SOC strongly alters the commonly known MI (immiscibility) condition, $%
g_{12}^{2}>g_{1}g_{2}$, for the binary superfluid with coefficients $g_{1,2}$
and $g_{12}$ of the intra- and inter-species repulsive interactions. In
fact, the binary BEC is always subject to the MI under the action of the
SOC, which implies that ground state of the system is plausibly represented
by a striped phase.
\end{abstract}

\maketitle


\section{Introduction}

Bose-Einstein Condensates (BECs) display a great variety of phenomena which
may be efficiently used for simulating diverse effects known in nonlinear
optics, condensed matter, and other physical settings \cite{simulator}. In
particular, a lot of interest has been recently drawn to the possibility of
implementing artificially engineered spin-orbit coupling (SOC) in the spinor
(two-component) BEC \cite{Linn,SOC}. SOC interactions, in the Dresselhaus
and Rashba \cite{Rashba} forms, account for a number of fundamental
phenomena in semiconductor physics, such as the spin-Hall effect \cite{Kato}
topological superconductivity \cite{Hasan}, and realization of spintronics
\cite{spintronics}. In solid-state settings, the SOC manifests itself in
lifting the degeneracy of single-electron energy levels by linking the spin
and orbital degrees of freedom. In the binary BEC, the synthetic SOC may be
induced via two-photon Raman transitions which couple two different
hyperfine states of the atom. In the first experiment \cite{Linn}, the pair
of states $\left\vert F=1,m_{F}=0\right\rangle $ and $\left\vert
F=1,m_{F}=-1\right\rangle $ of $^{87}$Rb atoms were used for this purpose.
In the combination with the intrinsic matter-wave nonlinearity, the SOC
setting offers a platform for the studies of various patterns and collective
excitations in the condensates. These studies address the
miscibility-immiscibility transition \cite{Gautam} and the structure and
stability of various nonlinear states, including specific structures of the
ground state \cite{Stanescu}, the Bloch spectrum in optical lattices \cite%
{Hamner}, Josephson tunneling \cite{Zhang}, fragmentation of condensates
\cite{Song}, tricritical points \cite{tricrit}, striped phases \cite{stripes}%
, supercurrents \cite{supercurrents}, vortices and vortex lattices \cite%
{vortices}, solitons, in one- \cite{solitons1D}, two- \cite{solitons2D} and
three- \cite{solitons3D} dimensional settings, optical and SOC states at
finite temperatures \cite{He} etc. Effects of SOC in degenerate Fermi gases
were considered too \cite{fermions}.

A fundamental ingredient of the matter-wave dynamics is the modulation
instability (MI) of flat (continuous-wave, CW) states against small
perturbations initiating the transformation of the constant-amplitude CW
into a state with a modulated amplitude profile. The MI, alias the
Benjamin-Fier instability \cite{BF}, is the key mechanism for the formation
of soliton trains in diverse physical media, as a result of the interplay
between the intrinsic nonlinearity and diffraction/dispersion (or the
kinetic-energy term, in terms of the matter-wave dynamics) \cite{Agrawal}.
The nonlinearity in ultracold atomic gases is induced by inter-atomic
collisions, which are controlled by the \textit{s}-wave scattering length.
The scattering lengths itself may be controlled by optical \cite{Blatt} and
magnetic \cite{Inouye} Feshbach resonances, while the SOC\ strength may be
adjusted as a function of the angle between the Raman laser beams, and their
intensity \cite{Higbie}.

The MI in single-component BECs has been addressed in many earlier works
\cite{Agrawal,Theochairs,Salasnich}. As in other physical systems \cite%
{BF,Agrawal}, it was concluded that the single-component MI is possible only
for the self-focusing (attractive) sign of the nonlinearity \cite%
{Higbie,Khawaja}. In attractive BECs, the formation of soliton trains is
initiated by phase fluctuations via the MI \cite{Strecker}. In two-component
BECs, the MI, first considered by Goldstein and Meystre \cite{Goldstein}, is
possible even for repulsive interactions \cite{Kasamatsu1,Kasamatsu2},
similar to the result known in nonlinear optics \cite{Agrawal-XPM}, in the
case when the XPM (cross--phase-modulation)-mediated repulsion between the
components is stronger than the SPM (self-repulsion) of each component. In
such a case, the MI creates not trains of bright solitons, but rather domain
walls which realize the phase separation in the immiscible binary BEC \cite%
{Poland,Kasamatsu1,Kasamatsu2,Ho,Vidanovic,Pu,Robins,Li1,Mithun}.

The objective of the present work is the analysis of the MI in the
effectively one-dimensional SOC system in the framework of the mean-field
approach. This is inspired, in particular, by the recent studies of the
dynamical instability of supercurrents, as a consequence of the violation of
the Galilean invariance by the SOC in one dimension (1D) \cite{supercurrents}%
, 2D instability at finite temperatures \cite{He}, and phase separation
under the action of the SOC \cite{Kasamatsu2,Gautam}. The character of the
MI, i.e., the dependence of its gain on the perturbation wavenumber, and the
structure of the respective perturbation eigenmodes, determine the character
of patterns to be generated by the MI. In particular, one may expect that
the MI will lead to the generation of striped structures which realize the
ground state in the SOC BEC with the repulsive intrinsic nonlinearity \cite%
{stripes}.

The subsequent material is structured as follows. In section II, we present
the model based on the system of coupled Gross-Pitaevskii equations (GPEs)
for the two-component BEC, including the SOC terms and collisional
nonlinearity. In this section, we also derive the dispersion relation for
the MI by means of the linear-stability analysis. Results of the systematic
analysis of the MI in the binary condensate are summarized in section III.
The paper is concluded by section IV.

\section{The model and modulational-instability (MI) analysis}

We take the single-particle Hamiltonian that accounts for the SOC of the
combined Dresselhaus-Rashba type, induced by the Raman lasers illuminating
the binary BEC. As is known, it can be cast in the form of \cite%
{Linn,Li,Cheng}
\begin{equation}
\hat{H}_{0}=\frac{{\hat{p}_{x}}^{2}}{2m}+\frac{\hbar k_{L}}{m}{\hat{p}%
_{x}\sigma _{z}}+\frac{\hbar \Omega }{2}{\sigma _{x}}+V(x),  \label{H0}
\end{equation}%
where ${\hat{p}}_{x}$ is the 1D momentum, $k_{L}$ is the SOC\ strength, $%
\Omega $ is the Rabi frequency of the linear mixing, $\sigma _{x,z}$ are the
Pauli matrices, and $V(x)$ is the trapping potential. Adding the collisional
nonlinear terms, the Hamiltonian produces the system of coupled GPEs for
scaled wave functions of the two components, $u_{1,2}=\psi _{1,2}/\sqrt{a}%
_{\bot }$, where $a_{\perp }\equiv \sqrt{\hbar /(m\omega _{\bot })}$, and $%
\omega _{\perp }$ is the radius of the transverse confinement which reduces
the 3D geometry to 1D \cite{Li,Cheng}:
\begin{gather}
i\frac{\partial u_{j}}{\partial t}=-\frac{1}{2}{\frac{\partial ^{2}u_{j}}{%
\partial x^{2}}}+i(-1)^{j}\gamma {\frac{\partial u_{j}}{\partial x}}+\Gamma
u_{3-j}+  \notag \\
(g_j|u_{j}|^{2}+g_{12}|u_{3-j}|^{2})u_{j}+V(x)u_{j},~j=1,2,  \label{GPE}
\end{gather}%
where the length, time, 1D atomic density, and energy are measured in units
of $a_{\bot }$, ${\omega _{\bot }}^{-1}$, ${a_{\bot }}^{-1}$ and $\hbar
\omega _{\bot }$, respectively. Further,\ scaled nonlinearity coefficients
are $g_j={2a_j}/{a_{\bot }}$ and $g_{12}={2a_{12}}/{a_{\bot }}$, where $a$ and $%
a_{12}$ are scattering lengths of the intra- and inter-component atomic
collisions. Finally, $\gamma \equiv k_{L}a_{\perp}$ and $\Gamma \equiv {\Omega }/%
{2\omega _{\bot }}$ are the scaled strengths of the SOC and Rabi coupling,
respectively, that may be defined to be positive, except for the situation
represented by Fig. \ref{fig10}, see below. The number of atoms in each
component is given by $N_{1,2}=\int_{-\infty }^{+\infty }\!|u_{1,2}|^{2}\,%
\mathrm{d}x$.

In the framework of Eq. (\ref{GPE}), we first address the MI of the CW state
in the form of a miscible binary condensate with uniform densities $n_{10}$
and $n_{20}$, and a common chemical potential, $\mu $, of both components: $%
u_{j}=e^{-i\mu t}\sqrt{n_{j0}}$. In the absence of the trapping potential,
the densities are determined by algebraic equations \cite%
{Goldstein,Kasamatsu1,Kasamatsu2}
\begin{equation}
\Gamma =-{\sqrt{\frac{n_{j0}}{n_{3-j,0}}}(g_jn_{j0}+g_{12}n_{3-j,0}-\mu )}%
,~j=1,2.  \label{nn}
\end{equation}%
For perturbed wave functions of the form $u_{j}=(\sqrt{n_{j0}}+\delta \psi
_{j})e^{-i\mu t}$, linearized equation for the small perturbations are
\begin{equation}
\begin{split}
i\frac{\partial \left( \delta \psi _{1}\right) }{\partial t}&=-\frac{1}{2}{%
\frac{\partial ^{2}\left( \delta \psi _{1}\right) }{\partial x^{2}}}-i\gamma
{\frac{\partial \left( \delta \psi _{1}\right) }{\partial x}}+ \Gamma (\delta \psi _{2}-\sqrt{\frac{n_{20}}{n_{10}}}\delta\psi_1) \\\
& 
+g_1n_{10}(\delta \psi _{1}+\delta {\psi _{1}}^{\ast })+g_{12}\sqrt{%
n_{10}n_{20}}(\delta \psi _{2}+\delta {\psi _{2}^{\ast }}),  
\end{split}
\end{equation}
\begin{equation}
\begin{split}
i\frac{\partial \left( \delta \psi _{2}\right) }{\partial t}&=-\frac{1}{2}{%
\frac{\partial ^{2}\left( \delta \psi _{2}\right) }{\partial x^{2}}}+i\gamma
{\frac{\partial \left( \delta \psi _{2}\right) }{\partial x}}+\Gamma (\delta
\psi _{1}-\sqrt{\frac{n_{10}}{n_{20}}}\delta\psi_2)\\
&+g_2n_{20}(\delta \psi _{2}+\delta {\psi _{2}}^{\ast })+g_{12}\sqrt{%
n_{10}n_{20}}(\delta \psi _{1}+\delta {\psi _{1}^{\ast }}),  \label{lin2}
\end{split}
\end{equation}%
where $\ast $ stands for the complex conjugate. We look for
eigenmodes of the perturbations in the form of plane waves, $\delta \psi
_{j}=\zeta _{j}\cos (kx-\Omega t)+i\eta _{j}\sin (kx-\Omega t)$, with real
wavenumber $k$ and, generally, complex eigenfrequency $\Omega $ and
amplitudes $\zeta _{j}$, $\eta _{j}$. The substitution of this in Eqs. (\ref%
{lin1}) and (\ref{lin2}) produces the dispersion relation for eigenfrequency
$\Omega $:
\begin{gather}
\Omega ^{4}+\Omega ^{2}\left[ -\frac{1}{4}{(k^{2}-2\Gamma
)(2k^{2}+G_{1}+G_{2})}-2k^{2}\gamma ^{2}-2\Gamma G_{12}\right]  \notag \\
+\frac{\Omega }{2}\left[ k\gamma (k^{2}-2\Gamma )(G_{2}-G_{1})\right]  \notag
\\
+k^{2} \left[ \gamma ^{2} \left(k^{2}\gamma ^{2}+2\Gamma G_{12}-\frac{1}{4}{%
(k^{2}-2\Gamma )(2k^{2}+G_{1}+G_{2})}\right.\right)  \notag \\
+\left( \frac{k^{2}}{4}-\Gamma\right) \left. \left( \frac{%
(k^{2}+G_{1})(k^{2}+G_{2})}{4}-G_{12}^{2}\right) \right] =0,  \label{disprel}
\end{gather}%
where we define
\begin{equation}
G_{1}\equiv 4g_1n_{10}-2\Gamma ,G_{2}\equiv 4g_2n_{10}-2\Gamma ,~G_{12}\equiv
2g_{12}n_{10}+\Gamma .\newline
\label{G}
\end{equation}

Quartic equation (\ref{disprel}) for $\Omega$ obtained for equal densities of the
two components, $n_{10}=n_{20}$ may be simplified to a practically solvable form by assuming that the strengths of the intraspecies interactions are equal, $G_{1}=G_{2}\equiv G$  [in this case, Eq. (\ref{nn}) determines the
corresponding chemical potential, $\mu $]. The result is
\begin{equation}
{\Omega _{\pm }^{2}}=\frac{1}{2}\left( \Lambda \pm \sqrt{\Lambda ^{2}+4R}%
\right) ,  \label{Omega}
\end{equation}%
\newline
where we define%
\begin{equation}
\Lambda =2k^{2}\gamma ^{2}+\frac{1}{2}{(k^{2}-2\Gamma )(k^{2}+G)}+2\Gamma
G_{12},
\end{equation}%
\begin{equation}
R=(S-\Lambda _{+}\Lambda _{-}),\newline
\end{equation}%
\begin{equation}
\Lambda _{\pm }=\frac{\Lambda \pm \sqrt{2\Lambda _{1}(\Lambda -\Lambda _{1})}%
}{2},
\end{equation}%
\begin{gather}
S=\frac{1}{4}\left\{ {(k^{2}-2\Gamma )^{2}}G_{12}^{2}+(G+k^{2})^{2}\Gamma
^{2}\right\}   \notag \\
+\frac{1}{2}\left\{ \gamma ^{2}k^{2}(G+k^{2})(k^{2}-2\Gamma )\right\} ,
\end{gather}%
\begin{equation}
\Lambda _{1}=\frac{1}{2}{(k^{2}-2\Gamma )(G+k^{2})}
\end{equation}%
\newline
The expression given by Eq. (\ref{Omega}) may be positive, negative or
complex, depending on the signs and magnitudes of the terms involved. The CW
state is stable provided that ${\Omega _{\pm }^{2}}>0$ for all real $k$;
otherwise, the instability growth rate is defined as $\xi \equiv \left\vert
\text{\textrm{Im}}\left( \Omega _{\pm }\right) \right\vert $. Thus, the MI
takes place, with complex ${\Omega _{\pm }^{2}}$, at ${\Lambda }^{2}+4R<0$.
At $R>0$, ${\Omega _{+}^{2}}$ is always positive for $\Lambda >0$, hence the
CW state is stable against the growth of perturbations accounted for by $%
\Omega _{+}$. At $\Lambda <0$, ${\Omega _{+}^{2}}$ is negative in the range
of $-{\Lambda }^{2}/4 \le  R<0$, where the CW state is unstable. Irrespective
of the value of $\Lambda $ but for $R>0$, the MI always sets in via the
growth of the perturbations which are accounted for by ${\Omega _{-}^{2}<0}$%
. The commonly known case of the MI in the single-component model, which
corresponds to $\gamma =\Gamma =0$, i.e., $G=4g$ and $G_{12}=0$ \cite%
{Agrawal,Theochairs,Salasnich}, is reproduced by the above results: it
occurs for $g<0$ (self-attractive nonlinearity) in the interval of
perturbation wavenumbers $0<k<2\sqrt{|g|}$, with the maximum gain, $\xi
\equiv \mathrm{Im}\left( \Omega \right) $, attained at $k_{\max }=2\sqrt{g}$.

\section{Results and discussions}

For the experimentally realized SOC system in the condensate of $^{87}$Rb
atoms, equally distributed between the two pseudo-spin states ($%
n_{10}=n_{20}\equiv n_{0}$; by means of rescaling, we set the common CW
density of both components to be $n_{0}=1$), the collisional nonlinearity is
repulsive, i.e., $g>0$ and $g_{12}>0$ in Eq. (\ref{GPE}). Here, for the sake
of generality, we are going to consider the MI in this case, as well as in
the system with other signs of the nonlinear terms. Then, depending on the
signs of $G$ and $G_{12}$, defined per Eq. (\ref{G}), four different cases
arise. Also, we will compare our results with previously obtained results in
absence of SOC.

\subsection{MI in the absence of SOC}

In the absence of SOC, we here dwell on two special cases, which are $\Gamma
=0$ and $\Gamma \neq 0$.

\subsubsection{Zero Rabi coupling}

For $\Gamma =0$, the model amounts to the usual two-component system, and
our solution for $\Omega ^{2}$ is
\begin{equation}
{\Omega _{\pm }^{2}}=\frac{k^{2}}{2}\left( \frac{k^{2}}{2}+2gn_{10}\pm
2g_{12}n_{10}\right) ,  \label{case1}
\end{equation}%
which exactly matches the one obtained in Refs. \cite{Kasamatsu1, Kasamatsu2}%
. It is well known that, in this case, for repulsive inter- and
intra-component interactions, the MI occurs only when $g_{12}^{2}>g_{1}g_{2}$
\cite{Mineev}. Here it further reduces to $g_{12}>g$.

\subsubsection{Non-zero Rabi coupling}

\begin{figure}[tbp]
\includegraphics[width=0.4 \textwidth]{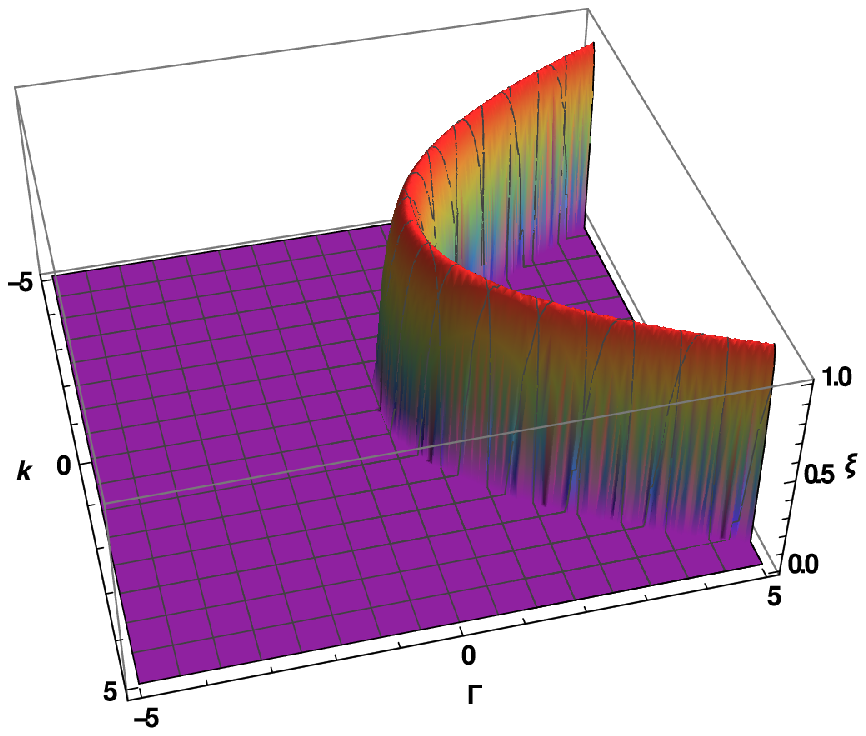}
\includegraphics[width=0.8cm, height=5cm]{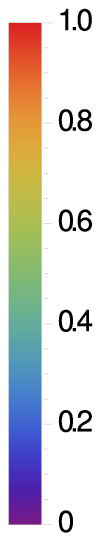}\newline
\caption{(Color online) The MI\ gain for $\Omega _{-}$ perturbations at $%
\protect\gamma =0$. Here $g=2$ and $g_{12}=1$. }
\label{fig1}
\end{figure}
In the presence of the Rabi coupling, Eq. \ref{Omega} is modified as
\begin{equation}
{\Omega _{+}^{2}}=\frac{k^{2}}{2}\left( \frac{k^{2}}{2}%
+2gn_{10}+2g_{12}n_{10}\right) ,  \label{case2ae}
\end{equation}%
\begin{equation}
{\Omega _{-}^{2}}=\frac{(k^{2}-4\Gamma )}{2}\left( \frac{k^{2}}{2}-2\Gamma
+2gn_{10}-2g_{12}n_{10}\right) .  \label{case2be}
\end{equation}%
The consideration of these solutions shows that $\Omega _{+}$ is real in the
following cases: i) both inter- and intra-component interactions are
repulsive, ii) attractive inter- and repulsive intra-component interactions,
only with $g\geq |g_{12}|$, and iii) repulsive inter- and attractive
intra-component interactions, only with $g\leq |g_{12}|$. In other cases, $%
\Omega _{+}$ is always imaginary. The effect of the Rabi coupling can be
seen from expression (\ref{case2be}) for $\Omega _{-}$. It is concluded
that, whenever $\Gamma $ is negative or zero, $\Omega _{-}$ is real for the
cases when i) both inter- and intra-component interactions are repulsive,
only with $g\geq |g_{12}|$, ii) attractive inter- and repulsive
intra-component interactions, and iii) attractive inter- and intra-component interactions, only with $|g|\leq |g_{12}|$. The situation is
similar to that for the case of zero Rabi coupling. The effect of the Rabi
coupling becomes appreciable at $\Gamma >0$. Figure \ref{fig1} shows the MI
gain when the strength of the inter-component interaction is smaller than
the intra-component strength, and it also shows the gain, $\xi =2\sqrt{%
(g-g_{12}-\Gamma )\Gamma }$, at $0<\Gamma <(g-g_{12})$. It is seen that the
usual MI (immiscibility) condition, $g_{12}>g$ , for the two-component BEC
system, with coefficients $g$ and $g_{12}$ of the intra- and inter-species
repulsive interactions, is not valid for $\Gamma >0$. On the other hand,
when the strengths of the intra- and inter-species repulsive interactions
are equal the gain is zero.

\subsection{$G>0$ and $G_{12}>0$}

\begin{figure}[tbp]
\centering
\includegraphics[width=0.4 \textwidth]{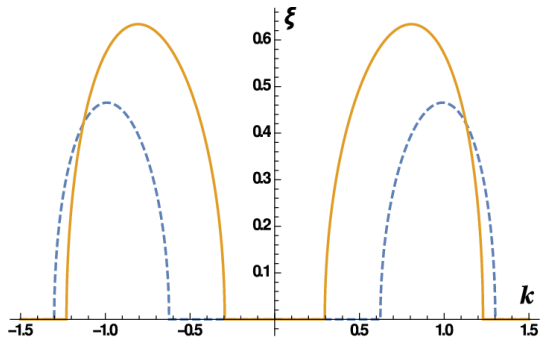}
\caption{(Color online) The gain of the MI against $\Omega _{+}$
perturbations for $\protect\gamma =\Gamma =1$. The solid line: $G=4~(g=1.5)$
and $G_{12}=3~(g_{12}=1)$. The dashed line: $G=2~(g=1)$ and $%
G_{12}=3~(g_{12}=1)$.}
\label{fig2}
\end{figure}

This case pertains to the case when both the intra- inter-component modified
interactions are repulsive. It follows from Eq. (\ref{Omega}) that, for $%
\Lambda >0$, expressions ${\Omega _{\pm }}$ are complex at ${\Lambda }%
^{2}+4R<0$. This corresponds to a single unstable region on the axis of the
perturbation wavenumber, $k$, as shown in Fig \ref{fig2}.

\begin{figure}[tbp]
\centering
\includegraphics[width=0.4\textwidth]{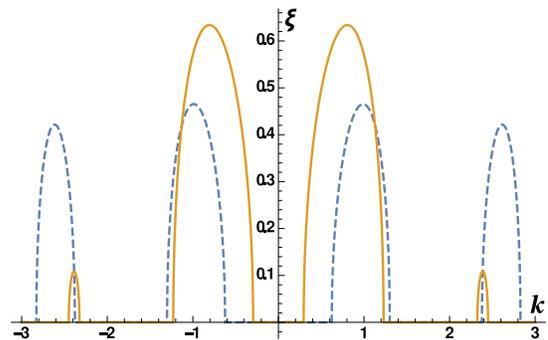}\newline
\caption{(Color online) The MI gain corresponding to the $\Omega _{-}$
perturbation branch for $\protect\gamma =\Gamma =1$. The solid line: $%
G=4~(g=1.5)$ and $G_{12}=3~(g_{12}=1)$. The dashed line: $G=2~(g=1)$ and $%
G_{12}=3~(g_{12}=1)$.}
\label{fig3}
\end{figure}

On the other hand, Eq. (\ref{Omega}) yields an unstable $\Omega _{-}$ branch
at ${\Lambda }^{2}+4R>0$ too. Two instability regions of $k$, corresponding
to this branch, are obtained, as shown in Fig. \ref{fig3}, for small values
of the interactions coefficient $G$ [see Eq. (\ref{G})], while the increase
of $G$ leads to merger of the two regions into one, as seen in Fig. \ref%
{fig4}, which shows the variation of the MI gain, $\xi $, and the range of
the values of wavenumber $k$ in which the MI holds, with the variation of $G$
and $G_{12}$. For smaller values of $G_{12}$ at fixed $G>0$, the MI\ gain in
the inner instability band gradually decreases to zero, while the gain in
the outer band at first vanishes, and then starts to grow with the growth of
$G_{12}$, as shown in Fig. \ref{fig5}. The increase in $G_{12}$ than $G$
makes term ${\Lambda }^{2}+4R$ positive, hence the MI occurs only at $\sqrt{{%
\Lambda }^{2}+4R}>0<\Lambda $, which shifts the gain band outwards. A
general conclusion of the above analysis is that, while, in the case of $%
0<g_{12}<g$, the usual system of the coupled GPEs does not give rise to the
MI, the inclusion of the SOC terms makes all the symmetric ($n_{10}=n_{20}$)
CW states, including those with $0<G_{12}<G$, modulationally unstable.

\begin{figure}[tbp]
\includegraphics[width=0.4\textwidth]{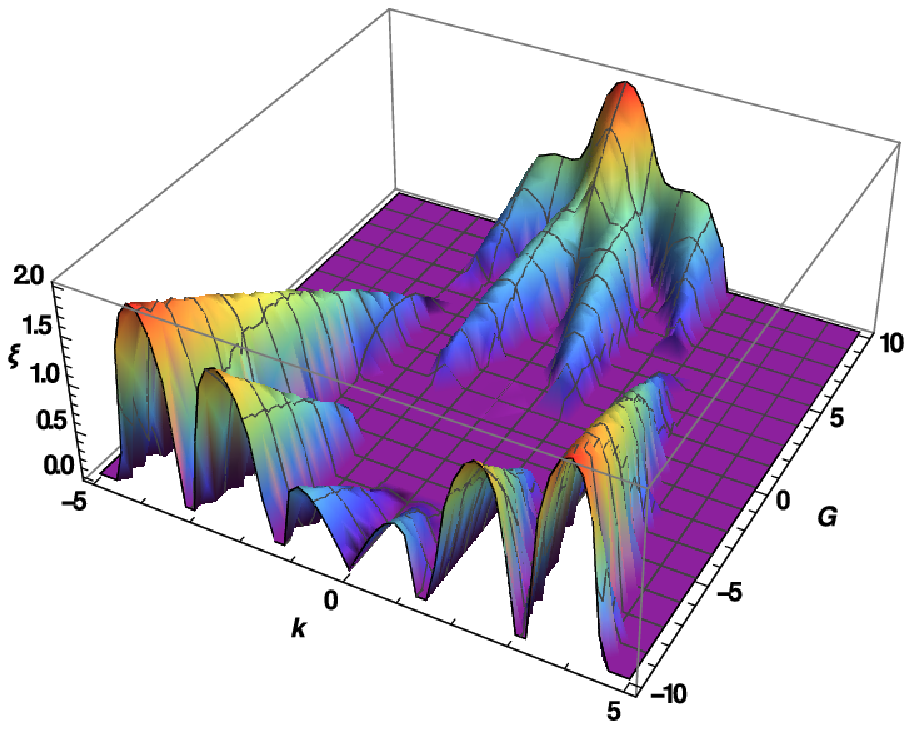}
\includegraphics[width=0.8cm, height=5cm]{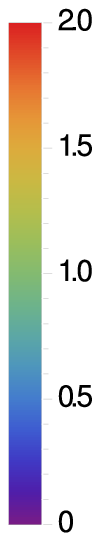}\newline
\caption{(Color online) The change of the MI\ gain with the variation of $G$
for fixed $G_{12}=3$ ($g_{12}=1$). Note that $G$ takes both positive and
negative values. Here $\protect\gamma =\Gamma =1$.}
\label{fig4}
\end{figure}
\begin{figure}[tbp]
\includegraphics[width=0.4\textwidth]{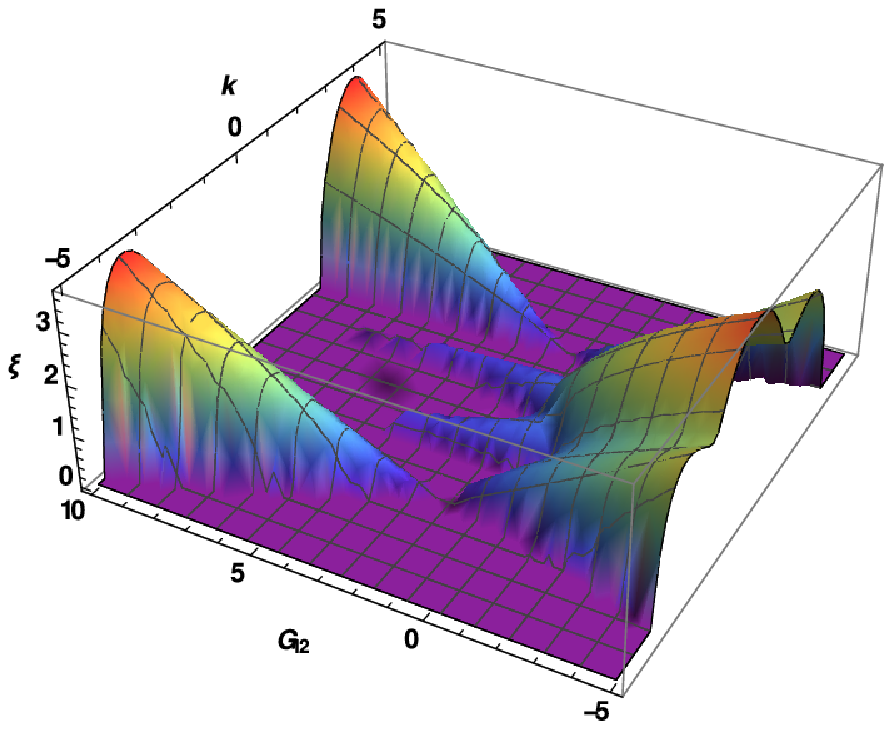}
\includegraphics[width=0.8cm, height=5cm]{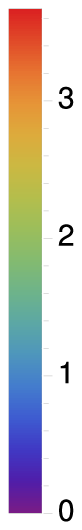}\newline
\caption{(Color online) The change of the MI gain with the the variation of $%
G_{12}$ for fixed $G=2$ ($g=1$). Note that $G_{12}$ takes both positive and
negative values. Here $\protect\gamma =\Gamma =1$.}
\label{fig5}
\end{figure}

\subsection{$G<0$ and $G_{12}>0$}

\begin{figure}[tbp]
\includegraphics[width=0.4\textwidth]{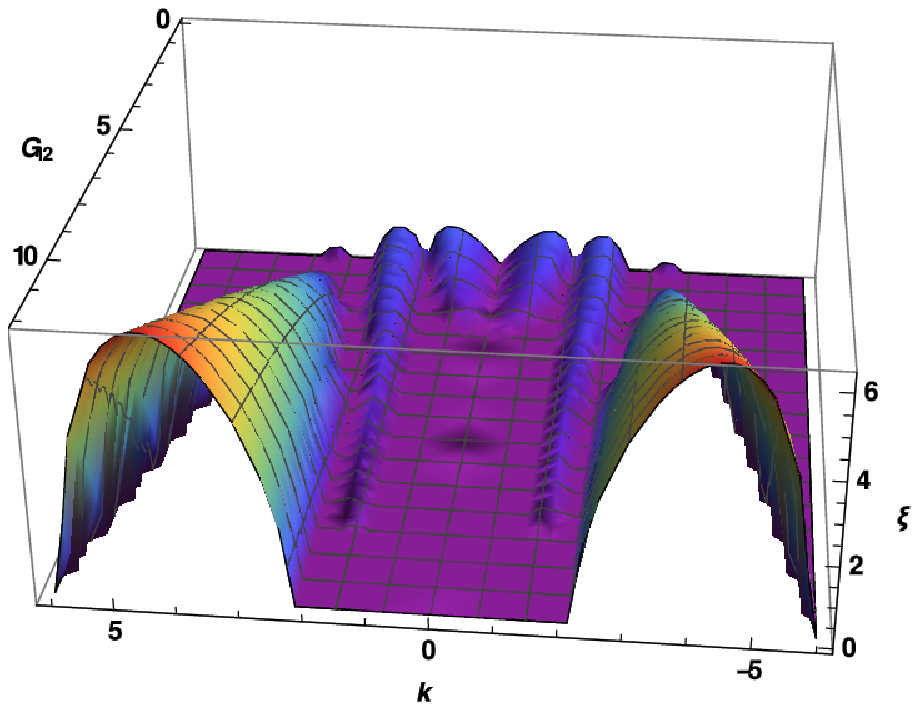}
\includegraphics[width=0.8cm, height=5cm]{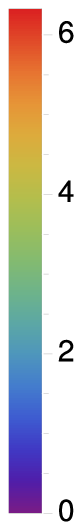}\newline
\caption{(Color online) The change of the MI gain with variation of $G_{12}$
at fixed $G=-6$ ($g=-1$). Here $\protect\gamma =\Gamma =1$.}
\label{fig6}
\end{figure}

This situation refers to the binary BEC with attractive intra-component and
repulsive inter-component interactions, which is subject to the MI in the
absence of the SOC, although the SOC may essentially affect the instability.
Here ${\Omega _{+}}$ produces the same number of MI bands as in the previous
subsection, while $\Omega _{-}$ produces more bands. For the $\Omega _{-}$
branch, the largest MI\ gain in the inner band is always smaller than in the
outer one, both increasing with the growth of $G$ at fixed $G_{12}$, as
shown in Fig. \ref{fig4}, while the MI\ gain in the inner MI band decreases
to zero, and increases in the outer band with the growth of $G_{12}$ at
fixed $G$, as in previous case, as shown in Fig. \ref{fig6}.

\subsection{$G>0$ and $G_{12}<0$}

In this case, the inter-component interaction is attractive, while the
intra-component nonlinearity is self-repulsive, and the MI of the CW with
equal densities of both components occurs provided that $\left\vert
g_{12}\right\vert >g$, in the absence of the SOC and Rabi coupling, while
the latter condition is not relevant for $\Gamma >0$ with $\gamma =0$, see
Eq. (\ref{case2be}). If the SOC terms are included, a single MI region is
generated by both $\Omega _{-}$ and $\Omega _{+}$ perturbation branches.
This is possible only for ${\Lambda }^{2}+4R>0$ with $\Lambda <0$. The MI
accounted for by $\Omega _{-}$ occurs in the range of $0\leq k<\sqrt{\beta
+2\beta _{1}+\sqrt{\beta ^{2}+4\beta \beta _{2}+4\beta _{1}^{2}}}$, where $%
\beta \equiv (2|G_{12}|-G)/2$, $\beta _{1}\equiv \gamma ^{2}+\Gamma $, $%
\beta _{2}\equiv \gamma ^{2}-\Gamma $, and the largest MI gain, $\xi _{\max
}=\sqrt{\Gamma (2|G_{12}|+G)}$, corresponds to $k=0$, as shown in Fig. \ref%
{fig7}.
\begin{figure}[tbp]
\centering
\includegraphics[width=0.4\textwidth]{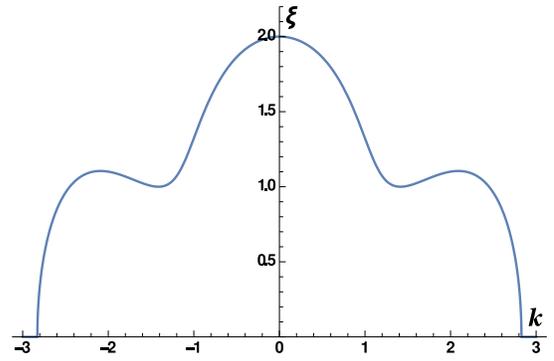}\newline
\caption{(Color online) The MI corresponding to the $\Omega _{-}$
perturbation branch with $\protect\gamma =\Gamma =1$, $G=2~(g=1)$ and $%
G_{12}=-1$ $(g_{12}=-1)$.}
\label{fig7}
\end{figure}

For a fixed value of $G$, which in this case is positive, local maxima of
the MI\ gain, $\xi $ (other than at $k=0$), gradually evolve with the growth
of $G_{12}$, resulting in diminished gain at $k=0$, as shown in Fig. \ref%
{fig8}.
\begin{figure}[tbp]
\centering
\includegraphics[width=0.4\textwidth]{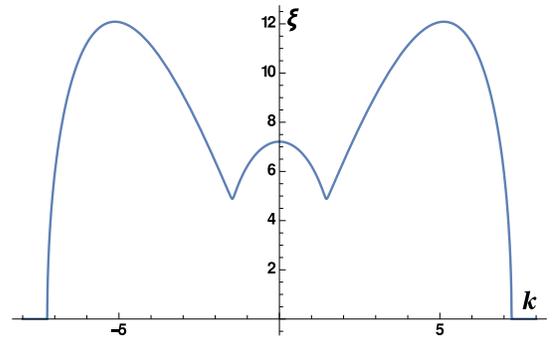}\newline
\caption{(Color online) The MI corresponding to the $\Omega _{-}$
perturbation branch with $\protect\gamma =\Gamma =1$, $G=2$ $(g=1)$ and $%
G_{12}=-25$ $(g_{12}=-13)$.}
\label{fig8}
\end{figure}

\subsection{$G < 0$ and $G_{12} < 0$}

\begin{figure}[tbp]
\includegraphics[width=0.4\textwidth]{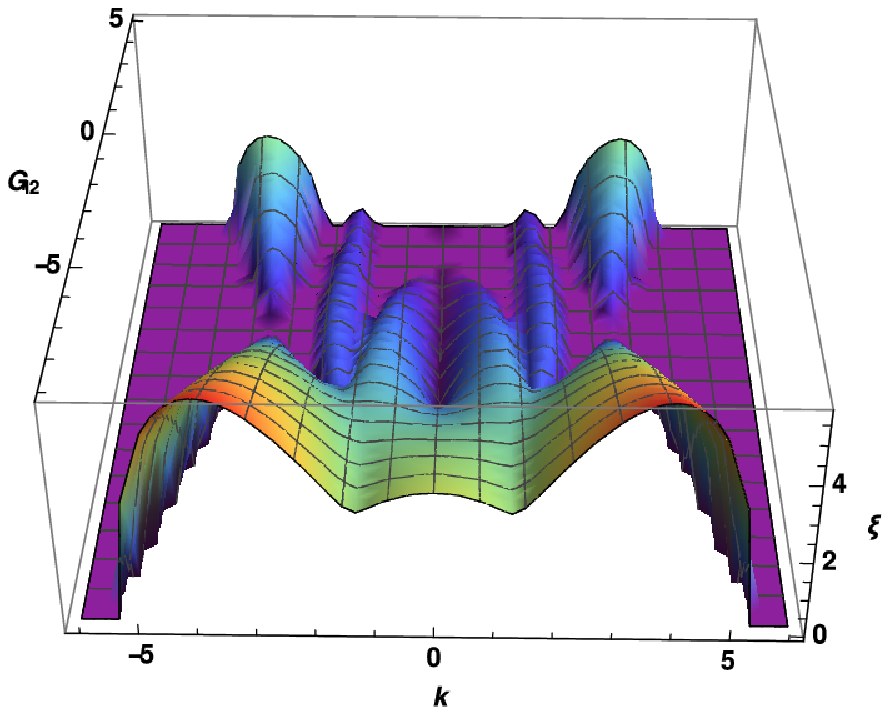}
\includegraphics[width=0.8cm, height=5cm]{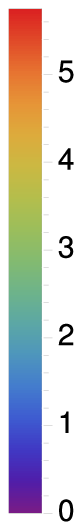}\newline
\caption{(Color online) The change of the MI\ gain with the variation of $%
G_{12}$ for fixed $G=-6$ ($g=-1$).}
\label{fig9}
\end{figure}

When both the intra- and inter-component interactions are attractive,
multiple MI\ bands are, quite naturally, formed for small values of $G_{12}$
at fixed $G<0$. They ultimately merge into a single band around $k=0$, as
shown in Fig. \ref{fig9}. In Figs. \ref{fig8} and \ref{fig9}, it is seen
that, with the growth of $G_{12}$, the situation for the present case
effectively simplifies into that reported in the previous subsection,
implying that the MI becomes independent of the nature of the
intra-component interaction for the strong inter-component attraction.

\subsection{Effect of the spin-orbit-coupling on the modulational instability%
}

\begin{figure}[tbp]
\includegraphics[width=0.4\textwidth]{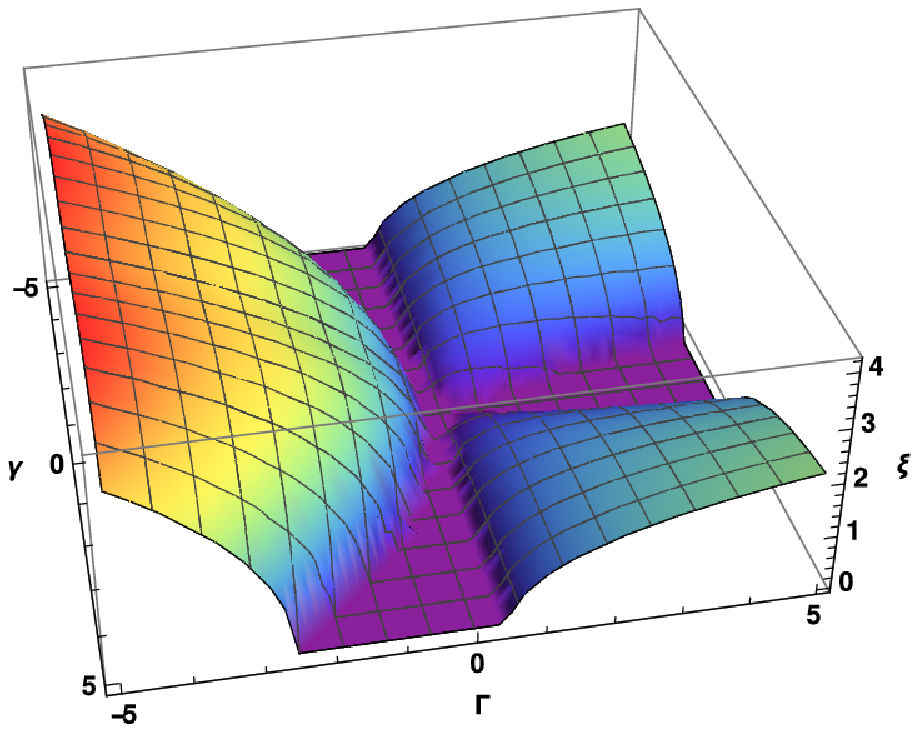}
\includegraphics[width=0.8cm, height=5cm]{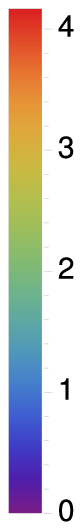}\newline
\caption{(Color online) The MI gain corresponding to the $\Omega _{-}$
perturbation branch for $G=2$,$G_{12}=3$ and $k=1$, as a function of the SOC
coefficient, $\protect\gamma $ and $\Gamma $.}
\label{fig10}
\includegraphics[width=0.4\textwidth]{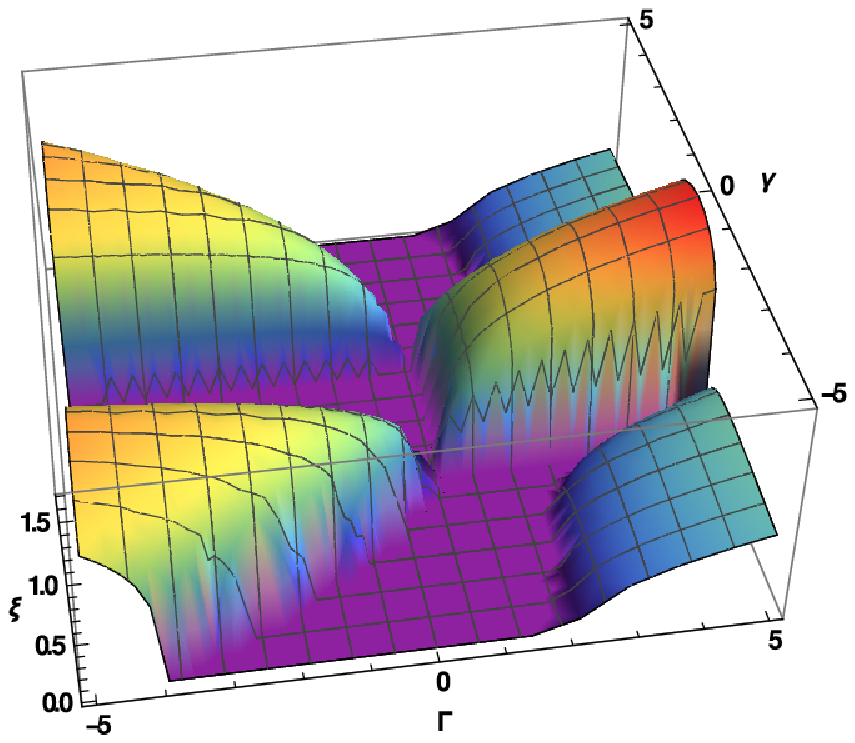}
\includegraphics[width=0.8cm, height=5cm]{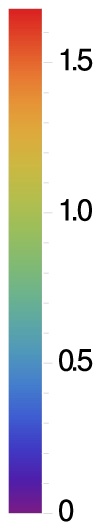}
\caption{(Color online) The MI gain corresponding to the $\Omega _{-}$
perturbation branch for $G=10$, $G_{12}=3$, and $k=2$, as a function the SOC
coefficients, $\protect\gamma $ and $\Gamma $,}
\label{fig11}
\end{figure}

The above results display the MI gain as a function of the modified
nonlinearity coefficients, $G$ and $G_{12}$, at fixed values of the SOC
parameters, $\gamma $ and $\Gamma $. It is relevant too to display the gain
as a function of the latter parameters, for a fixed strength of the
nonlinearity. To this end, Fig. \ref{fig10} shows the MI gain versus $\gamma
$ and $\Gamma $, for fixed $G$ and $G_{12}$. The fixed value of the
perturbation wavenumber, $k=1$, is chosen from Fig. \ref{fig3}, where the
gain's maximum is observed for $G=2$ and $G_{12}=3$ at $k=1$.

As mentioned above, it is commonly known that, in the two-component
repulsive condensates, the MI, leading to the immiscibility of the binary
superfluid, occurs in the region of $g_{12}^{2}>g_{1}g_{2}$ (if the
self-repulsion coefficients are different in the two components, $g_{1}\neq
g_{2})$ \cite{Mineev,Timmermans,Kasamatsu1,Kasamatsu2}. On the other hand,
it was more recently demonstrated that the linear interconversion between
the components (accounted for by coefficient $\Gamma $ in the present
notation) shifts the miscibility threshold to larger values of $g_{12}$ \cite%
{Merhasin}. In the presence of the SOC, Figs. \ref{fig2} and \ref{fig3}
clearly show that the MI also occurs at $g_{12}\leq g$, for $g_{1}=g_{2}$
and equal CW densities in the two components.

In the latter connection, it is relevant to mention that, for the fixed
values $G=2$ and $G_{12}=3$, and setting $\gamma =0$ (no SOC proper, while
the Rabi mixing is present, $\Gamma \neq 0$), the MI condition, $g_{12}>g$,
holds in the range\ of $|\Gamma |<1$. Figure \ref{fig10} shows that the MI
is indeed present in this range at $\gamma =0$. On the other hand, the same
figure shows zero MI gain for $g_{12}\leq g$ and $\gamma =0$. This
observation elucidates the validity of our analysis. In order to see the
effect of $\gamma $, we increase the strength of intra-component interaction
to $G=10$. For this value, condition $g_{12}<g$ holds for the region of $%
-1<\Gamma <3$. Figure \ref{fig11} shows that the MI gain does not vanish in
this region. This shows that, in presence of the SOC, irrespective of the
sign of the Rabi coupling, the system with repulsive interactions is always
subject to the MI.

The results of the analysis of the MI are summarized in Table I.

\begin{table*}[!btp]
\caption{Summary of results the MI analysis obtained for different
combination of the parameters. }{\footnotesize \label{tab1}  \centering
\begin{tabular}{p{2cm}p{1cm}p{1cm}p{1cm}p{1cm}p{9cm}c}
\hline\hline
Cases & $\gamma$ & $\Gamma$ & $G$ & $G_{12}$ & Inference &  \\[0.5ex] \hline
\multirow{2}{*}{1} & \multirow{2}{*}{0} & \multirow{2}{*}{0} & + & %
\multirow{2}{*}{0} & Always Stable &  \\
&  &  & - &  & Always unstable &  \\ \hline
\multirow{4}{*}{2} & \multirow{4}{*}{0} & \multirow{4}{*}{0} & + & + &
Unstable only if $g_{12}>g$ &  \\
&  &  & + & - & Unstable only if $|g_{12}|>g$ &  \\
&  &  & - & + & \multirow{2}{*}{Always unstable} &  \\
&  &  & - & - &  &  \\ \hline
\multirow{4}{*}{3} & \multirow{4}{*}{0} & \multirow{4}{*}{$\ne 0$} & + & + & %
\multirow{4}{*}{\shortstack{For $\Gamma>0$, always unstable, except for the
case of $g=g_{12}$ when \\ both $G$ and $G_{12}$ are positive, and
miscibility condition $g_{12}>g$ is \\ invalid. For $\Gamma \le 0$, this
special case reduces to case 2}} &  \\
&  &  & + & - &  &  \\
&  &  & - & + &  &  \\
&  &  & - & - &  &  \\ \hline
\multirow{4}{*}{4} & \multirow{4}{*}{$\ne 0$} & \multirow{4}{*}{$\ne 0$} & +
& + & \multirow{4}{*}{\shortstack{The binary BEC is always vulnerable to the MI,
irrespective of \\ the nature of the nonlinear interactions }} &  \\
&  &  & + & - &  &  \\
&  &  & - & + &  &  \\
&  &  & - & - &  &  \\[1ex] \hline\hline
\end{tabular}
}
\end{table*}

\section{Conclusions}

We have considered the modulational instability (MI) of the flat CW
background with equal densities of two components in the nonlinear BEC,
subject to the action of the spin-orbit coupling (SOC). This was done by
means of the linear-stability analysis for small perturbations added to the
CW states. The SOC acting on the binary BEC affects the phase separation or
mixing between the components, driven by the competition of intra- and
inter-component interactions. The effective modification induced by the SOC
is different for these interactions, the former and latter interactions
being, respectively, reduced and enhanced by the Rabi coupling [measured by
coefficient $\Gamma $ in Eq. (\ref{GPE})]. Our analysis shows that,
irrespective of the sign of inter and intra-component interactions, the
system is vulnerable to the MI. For the fixed strength of $\gamma $ and $%
\Gamma $, the MI becomes independent of the nature of the intra-component
interactions when the inter-component attraction is strong. The main result
of our analysis is the strong change of the commonly known phase-separation
condition $g_{12}^{2}>g_{1}g_{2}$ for the binary superfluid with repulsive
interactions.

It may be expected that the nonlinear development of the MI will lead to
establishment of multidomain patterns, such as the above-mentioned striped
ones (see, e.g., Ref. \cite{stripes}). Systematic numerical analysis of such
scenarios will be reported elsewhere.

\section{Acknowledgements}

K.P. thanks agencies DST, CSIR, NBHM, IFCPAR and DST-FCT, funded by the
Government of India, for the financial support through major projects.


\begin{thebibliography}{99}
\bibitem{simulator} P. Hauke, F. M. Cucchietti, L. Tagliacozzo, I. Deutsch,
and M. Lewenstein, Rep. Prog. Phys. \textbf{75}, 082401 (2012).

\bibitem{Linn} Y. J. Lin, K. Jimenez-Garcia, and I. B. Spielman, Nature
(London) \textbf{471}, 83(2011).

\bibitem{SOC} J. Dalibard, F. Gerbier, G. J\={u}zeliunas, and P. \"{O}hberg,
Rev. Mod. Phys. \textbf{83}, 1523 (2011); I. B. Spielman, Ann. Rev. Cold At.
Mol. \textbf{1}, 145 (2012); H. Zhai, Int. J. Mod. Phys. B \textbf{26},
1230001 (2012); V. Galitski and I. B. Spielman, Nature \textbf{494}, 49
(2013); X. Zhou, Y. Li, Z. Cai, and C. Wu, J. Phys. B: At. Mol. Opt. Phys.
\textbf{46}, 134001 (2013); N. Goldman, G. J\={u}zeliunas, P. \"{O}hberg,
and I. B. Spielman, Rep. Progr. Phys. \textbf{77}, 126401 (2014).

\bibitem{Rashba} G. Dresselhaus, Phys. Rev. \textbf{100}, 580 (1955); Y. A.
Bychkov and E. I. Rashba, J. Phys. C \textbf{17}, 6039 (1984).

\bibitem{Kato} Y. K. Kato, R. C. Myers, A. C. Gossard, and D. D. Awschalom,
Science\textbf{\ 306}, 1910 (2004).

\bibitem{Hasan} M. Z. Hasan and C. L. Kane, Rev. Mod. Phys. \textbf{82},
3045 (2010).

\bibitem{spintronics} I. \v{Z}uti\'{c}, J. Fabian, and S. Das Sarma, Rev.
Mod. Phys. \textbf{76}, 323 (2004); C. H. L. Quay, T. L. Hughes, J. A.
Sulpizio, L. N. Pfeiffer, K. W. Baldwin, K. W. West, D. Goldhaber-Gordon,
and R. de Picciotto, Nature Phys. \textbf{6}, 336 (2010).

\bibitem{Gautam} S. Gautam and S. K. Adhikari, Phys. Rev. A \textbf{90},
043619 (2014).

\bibitem{Stanescu} T. D. Stanescu, B. Anderson, and V. Galitski, Phys. Rev.
A \textbf{78}, 023616 (2008); S. Gopalakrishnan, A. Lamacraft, and P. M.
Goldbart, Phys. Rev. A \textbf{84}, 061604(R) (2011); M. Merkl, A. Jacob, F.
E. Zimmer, P. \"{O}hberg, and L. Santos, Phys. Rev. Lett. \textbf{104},
073603 (2010).

\bibitem{Hamner} C. Hamner, Y. Zhang, M. A. Khamehchi, M. J. Davis, and P.
Engels, Phys. Rev. Lett. \textbf{114}, 070401 (2015); Y. Zhang, and C.
Zhang, Phys. Rev. A \textbf{87}, 023611 (2013).

\bibitem{Zhang} D. W. Zhang, L. B. Fu, Z. D. Wang, and S.-L. Zhu, Phys. Rev.
A \textbf{85}, 043609 (2012).

\bibitem{Song} S. W. Song, Y. C. Zhang, H. Zhao, X. Wang, and W. M. Liu,
Phys. Rev. A \textbf{89}, 063613 (2014).

\bibitem{tricrit} Y. Li, L. P. Pitaevskii, and S. Stringari, Phys. Rev.
Lett. \textbf{108}, 225301 (2012).

\bibitem{stripes} C. Wang, C. Gao, C. M. Jian, and H. Zhai, Phys. Rev. Lett.
\textbf{105}, 160403 (2010); D. A. Zezyulin, R. Driben, V. V. Konotop, and
B. A. Malomed, Phys. Rev. A \textbf{88}, 013607 (2013).

\bibitem{supercurrents} T. Ozawa, L. P. Pitaevskii and S. Stringari, Phys.
Rev. A \textbf{87}, 063610 (2013).

\bibitem{vortices} B. Ramachandhran, B. Opanchuk, X.-J. Liu, H. Pu, P. D.
Drummond, and H. Hu, Phys. Rev. A \textbf{85}, 023606 (2012); H. Sakaguchi
and B. Li, \textit{ibid}. \textbf{87}, 015602 (2013); Y. Xu, Y. Zhang, and
B. Wu, \textit{ibid}. \textbf{87}, 013614 (2013); A. L. Fetter, \textit{ibid}%
. \textbf{89}, 023629 (2014)\thinspace ; P. Nikoli\'{c}, \textit{ibid}.
\textbf{90}, 023623 (2014).

\bibitem{solitons1D} V. Achilleos, D. J. Frantzeskakis, P. G. Kevrekidis,
and D. E. Pelinovsky, Phys. Rev. Lett. \textbf{110}, 264101 (2013); Y. V.
Kartashov, V. V. Konotop, and F. K. Abdullaev, Phys. Rev. Lett. \textbf{111}%
, 060402 (2013); Y. Xu, Y. Zhang, and B. Wu, Phys. Rev. A \textbf{87},
013614 (2013); L. Salasnich and B. A. Malomed, \textit{ibid}. \textbf{87}
063625 (2013); Y.-K. Liu and S.-J. Yang, EPL \textbf{108}, 30004 (2014); Y.
Cheng, G. Tang, and S. K. Adhikari, Phys. Rev. A \textbf{89}, 063602 (2014);
Y. V. Kartashov, V. V. Konotop, and D. A. Zezyulin, \textit{ibid}. \textbf{90%
}, 063621 (2014); Y. Zhang, Y. Xu, and T. Busch, \textit{ibid}. \textbf{91},
043629 (2015); P. Beli\v{c}ev, G. Gligori\'{c}, J. Petrovi\'{c}, A.
Maluckov, L. Had\v{z}ievski, and B. Malomed, J. Phys. B:\ At. Mol. Opt.
Phys. \textbf{48}, 065301 (2015); M. Salerno and F. Kh. Abdullaev, Phys.
Lett. A \textbf{379}, 2252 (2015); S. Gautam and S. K. Adhikari, Phys. Rev.
A \textbf{91}, 063617 (2015).

\bibitem{solitons2D} H. Sakaguchi, B. Li, and B. A. Malomed, Phys. Rev. E
\textbf{89}, 032920 (2014); H. Sakaguchi and B. A. Malomed, \textit{ibid}.
\textbf{90}, 062922 (2014); V. E. Lobanov, Y. V. Kartashov, and V. V.
Konotop, Phys. Rev. Lett. \textbf{112}, 180403 (2014); L. Salasnich, W. B.
Cardoso, and B. A. Malomed, Phys. Rev. A \textbf{90}, 033629 (2014); Y. Xu,
Y. Zhang, and C. Zhang, \textit{ibid}. \textbf{92}, 013633 (2015).

\bibitem{solitons3D} Y. -C. Zhang, Z. -W. Zhou. B. A. Malomed, and H. Pu,
arXiv:1509.04087.

\bibitem{He} P. S. He, J. Zhao, A. C. Geng, D. H. Xu, R. Hu, Phys. Lett. A
\textbf{377}, 2207 (2013).

\bibitem{fermions} P. Wang, Z. Q. Yu, Z. Fu, J. Miao, L. Huang, S. Chai, H.
Zhai, and J. Zhang, Phys. Rev. Lett. \textbf{109}, 095301 (2012); L. W.
Cheuk, A. T. Sommer, Z. Had\v{z}ibabi\'{c}, T. Yefsah, W. S. Bakr, and M. W.
Zwierlein, \textit{ibid}. \textbf{109}, 095302 (2012).

\bibitem{BF} T. B. Benjamin and J. E. Feir, J. Fluid Mech. \textbf{27}, 417
(1967)

\bibitem{Blatt} S. Blatt, T. L. Nicholson, B. J. Bloom, J. R. Williams, J.
W. Thomsen, P. S. Julienne, and J. Ye, Phys. Rev. Lett. \textbf{107}, 073202
(2011); M. Theis, G. Thalhammer, K. Winkler, M. Hellwig, G. Ruff, R. Grimm,
and J. H. Denschlag, \textit{ibid}. \textbf{93}, 123001 (2004); M. Yan, B.
J. DeSalvo, B. Ramachandhran, H. Pu, and T. C. Killian, \textit{ibid. }%
\textbf{110}, 123201 (2013).

\bibitem{Inouye} S. Inouye, M. R. Andrews, J. Stenger, H.-J. Miesner, D. M.
Stamper-Kurn, and W. Ketterle, Nature (London) \textbf{392}, 151 (1998); S.
L. Cornish, N. R. Claussen, J. L. Roberts, E. A. Cornell, and C. E. Wieman,
Phys. Rev. Lett. \textbf{85}, 1795 (2000).

\bibitem{Higbie} J. Higbie, and D. M. Stamper-Kurn, Phys. Rev. Lett. \textbf{%
88}, 090401(2002).

\bibitem{Agrawal} G. P. Agrawal,\textit{\ Non-linear Fiber Optics}, 5th ed.
(San Diego: Academic, 2013).

\bibitem{Khawaja} U. Al Khawaja, H. T. C. Stoof, R. G. Hulet, K. E.
Strecker, and G. B. Partridge, Phys. Rev. Lett. \textbf{89}, 200404 (2002).

\bibitem{Theochairs} G. Theocharis, Z. Rapti, P. G. Kevrekidis, D. J.
Frantzeskakis, and V. V. Konotop, Phys. Rev. A \textbf{67}, 063610 (2003)

\bibitem{Salasnich} L. Salasnich, A. Parola, and L. Reatto, Phys. Rev. Lett.
\textbf{\ 91}, 080405 (2003).

\bibitem{Strecker} K. E. Strecker, G. B, Partridge, A. G. Truscott, and R.
G. Hulet, Nature (London) \textbf{417}, 150 (2002); L. D. Carr and J. Brand,
Phys. Rev. Lett. \textbf{92} 040401 (2004); U. Al Khawaja, H. T. C. Stoof,
R. G. Hulet, K. E. Strecker, and G. B. Partridge, Phys. Rev. Lett. \textbf{89%
} 200404 (2002).

\bibitem{Goldstein} E. V. Goldstein, and P. Meystre, Phys. Rev. A \textbf{55}%
, 2935 (1997).

\bibitem{Kasamatsu1} K. Kasamatsu, and M. Tsubota, Phys. Rev. Lett. \textbf{%
93}, 100402 (2004).

\bibitem{Kasamatsu2} K. Kasamatsu, and M. Tsubota, Phys. Rev. A \textbf{74},
013617 (2006).

\bibitem{Agrawal-XPM} M. Yu, C. J. McKinstrie, and G. P. Agrawal, Phys. Rev.
E \textbf{48}, 2178-2186 (1993).

\bibitem{Poland} M. Trippenbach, K. Goral, K. Rzazewski, B. Malomed, and Y.
B. Band, J. Phys. B: At. Mol. Opt. \textbf{33}, 4017 (2000).

\bibitem{Ho} T.-L. Ho and V. B. Shenoy Phys. Rev. Lett. \textbf{77}. 3276
(1996)

\bibitem{Vidanovic} I. Vidanovi\'{c}, N. J. V. Druten and M. Haque, New J.
Phys. \textbf{15}, 035008 (2013)

\bibitem{Pu} H. Pu, C. K. Law, S. Raghavan, J. H. Eberly, and N. P. Bigelow,
Phys. Rev. A \textbf{60}, 1463 (1999).

\bibitem{Robins} N. P. Robins, W. Zhang, E. A. Ostrovskaya, and Y. S.
Kivshar, Phys. Rev. A \textbf{64}, 021601 (2001).

\bibitem{Li1} L. Li, Z. Li, B. A. Malomed, D. Mihalache, and W. M. Liu,
Phys. Rev. A \textbf{72}, 033611 (2005).

\bibitem{Mithun} T. Mithun and K. Porsezian, Phys. Rev. A \textbf{85},
013616 (2012).

\bibitem{Li} Y. Li, G. I. Martone, L. P. Pitaevskii, and S. Stringari, Phys.
Rev. Lett. \textbf{110}, 235302 (2013); L. Salasnich and B. A. Malomed,
Phys. Rev. A \textbf{87}, 063625 (2013).

\bibitem{Cheng} Y. Cheng, G. Tang, S. K. Adhikari, Phys. Rev. A, \textbf{89}%
, 063602 (2014).

\bibitem{Mineev} V. P. Mineev, Zh. Eksp. Teor. Fiz. \textbf{67}, 263 (1974)
[Sov. Phys. JETP \textbf{40}, 132 (1974)].

\bibitem{Merhasin} M. I. Merhasin, B. A. Malomed, and R. Driben, J. Phys.
B:\ At. Mol. Opt. Phys. \textbf{38}, 877 (2005).

\bibitem{Timmermans} E. Timmermans, Phys. Rev. Lett. \textbf{81}, 5718
(1998).
\end{thebibliography}
\end{document}